\documentclass[10pt]{article}
\usepackage[dvips]{graphicx}

\usepackage{epsfig}
\usepackage{graphicx}
\usepackage{indentfirst}
\usepackage{amsmath}
\usepackage{amsfonts}
\usepackage{amssymb}

\begin{document}
\begin{center}
{\Large\bf  Isotropic turbulence in the dark fluid universe with inhomogeneous equation of state \\}
\medskip

R. D. Boko$^{(a)}$ \footnote{e-mail:
docenzi@yahoo.fr}, M. E. Rodrigues$^{(b,c)}$\footnote{e-mail: esialg@gmail.com}, M. J. S. Houndjo$^{(a,d)}$\footnote{e-mail:
sthoundjo@yahoo.fr},  J. B. Chabi Orou$^{(a)}$\footnote{E-mail: jean.chabi@imsp-uac.org} and R. Myrzakulov$^{(e)}$\footnote{e-mail: rmyrzakulov@csufresno.edu}

$^{a}$ \,{\it Institut de Math\'{e}matiques et de Sciences Physiques (IMSP)} {\it 01 BP 613 Porto-Novo, B\'{e}nin}\\
$^{b}$ \,{\it Faculdade de F\'isica, PPGF, Universidade Federal do Par\'a,
66075-110, Bel\'em, Par\'a, Brazil}\\
$^{c}$\, {\it  Faculdade de Ci\^encias Exatas e Tecnologia, Universidade Federal do Par\'a - \\
Campus Universit\'ario de Abaetetuba, CEP 68440-000, Abaetetuba, Par\'a, Brazil}\\
$^{d}$\, {\it Facult\'e des Sciences et Techniques de Natitingou - Universit\'e de Parakou - B\'enin}\\
$^{e}$\, {\it Eurasian International Center for Theoretical Physics\\
L.N. Gumilyov Eurasian National University, Astana 010008, Kazakhstan}

\date{}

\end{center}
\begin{abstract}
We investigate the turbulence effect in dark fluid universe with linear inhomogeneous equation of state. Attention is attached to two physical situations. First, we perform the perturbative analysis of turbulence  and check its effects around the Big Rip. Later, treating the turbulence energy density as a part of total dark fluid, we study the stability of the system. The result shows that the stability is achieving as the energy density of turbulence decreases, changing into heat (the radiation), in perfect agreement with the avoidance of the Big Rip.
\end{abstract}

Pacs numbers: 98.80.-k,04.50.+h,11.10.Wx

\section{Introduction}
There is a lot of interest in the study of the nature of dark energy, responsible for the acceleration of the cosmic expansion \cite{1B1,4B1}, and which is expected to have  strange properties, like negative pressure and violates the strong energy condition. The current acceleration of the universe is realized when the parameter of equation of state (EoS) $\omega$, is such that $\omega<-1/3$. More precisely, $-1<\omega<-1/3$ corresponds to quintessence region where the expansion of the universe is accelerated, $\ddot{a}$, but not super-accelerated $\dot{H}>0$. When $\omega<-1$, the universe is in a phantom region, which usually ends by finite-time future singularities \cite{1B3}-\cite{19B3}.\par
Recently Brevik et al analyse the possibility of whether there can be a turbulence  micro-structure superimposed on the dark fluid \cite{B3}. In such situation, in order to maintain the macroscopic isotropic, attention is attached to the isotropic version of turbulence. An important aspect of this issue is that turbulence implies a loss of kinetic energy, changing it into heat. At macroscopic level, this turbulence is often treated as bulk viscosity. So, at microscopic level, within turbulence consideration, the length scale associated  with the macroscopic viscosity may be replaced by the Kolmogorov microlength scale, conventionally denoted by $\eta$ \cite{B3}. In that paper, authors assumed homogeneous equation of state and studied the influence of the turbulence on the evolution of the universe, mainly, in it dark sector.\par
In this paper, the goal remains the same, of studying the effects of the turbulence, but assuming inhomogeneous equation of state. We assume the EoS $p=\omega(t)\rho+\Lambda(t)$ in which $\omega(t)$ and $\Lambda(t)$ depend linearly on time. We also consider in this paper an oscillating $\omega(t)$. This particular king of EoS is one alternative amongst a variety of possibilities, proposed to cope with the general dark energy problem \cite{5B2,6B2}, and developed for various purposes in \cite{7B2}-\cite{14B2}.\par
Following the same steps of \cite{B3} we include the turbulence effect by adding a constant {\it fraction}, denoted $f$, to the laminar ordinary energy density in the first equation of FRW. Then, we perform a perturbative analysis to the first order, assuming $f$ to be a small quantity and the correction term to $\omega(t)$ and $\Lambda(t)$ have been found. Other interesting feature of considering this inhomogeneous EoS is that it generalizes the Brevik's results \cite{B3}, at least, when FRW equations with  the fraction $f$ of the energy as turbulence energy are assumed.\par
Moreover, instead of only considering the above perturbative treatment,  we also consider the total dark energy contribution as a laminar part plus a turbulence energy part. Then, we study the stability of the system. The existence of an attractor critical point is investigated which should correspond to the stability of the system. Within the assumption that the dynamics of the system  essentially depends on the evolution of the turbulence and the radiation, we find that stability corresponds to a decreasing turbulence energy density and an increasing radiation energy density. \par

The paper is organized as follows: in section $2$, we present the Kolmogorov's isotropic turbulence within inhomogeneous EoS. Section $3$ is devoted to the study of turbulence effects as a fraction of the energy in the FRW equations, and this, for various assumptions of parameters $\omega(t)$ and $\Lambda(t)$. We investigate the stability of the system in the section $4$ where turbulence is assumed as an additive  component to the total dark fluid. We present the conclusion in section $5$.

\section{Kolmogorov's isotropic turbulence within inhomogeneous equation of state}

A most effect of the turbulence is the loss of kinetic energy into heat. As it is shown in  
\cite{B3,25B3,26B3}, the macroscopic bulk viscosity may be substituted by the microscopic shear, from which the kolmogorov is directly proportional to $\nu^{3/4}$ and inversely proportional to $\epsilon^{1/4}$ where $\nu$ means the kinematic microscopic shear viscosity and $\epsilon$ the dissipation per unit of mass. By considering $l$ the external scale of the turbulence, whereby $1/l$ denotes the  corresponding wave number and assuming that the large eddies  move with a little dissipation of energy. According to the second hypothesis of Kolmogorov, in an isotropic region the motion is entirely by the friction and inertia \cite{B3,25B3}. Therefore, there is a continuous flux of energy transferred by the dissipation $\epsilon$. By denoting the size of an eddy by $\lambda$ and $k=1/\lambda$ the corresponding wave number. The equilibrium range for the wave number for which all the flow is lost is $k\gg 1/l$.\par
Let $u_{\lambda}$ characterizes the typical velocity of an eddy of size $\lambda$. Thus, the internal Reynolds number gets the form  Re${\lambda} \sim \lambda u_{\lambda}/\nu$.
For increasing values of $k$, Re${\lambda}$ decreases. When Re${\lambda}\sim 1$,  dissipation becomes important, corresponding to the condition which implies the Kolmogorov length. For large values of Reynolds number the quantities $1/l$ and $1/\eta$ are completely separated and there exists an inertial subrange characterized by $\frac{1}{l} \ll k \ll \frac{1}{\eta}$, 
where the fluid behaves like a non-viscous fluid \cite{B3}. In the inertial subrange the formula of the spectral energy density is $E(k)=\alpha \epsilon^{2/3}\,k^{-5/3}$
with $\alpha \approx 1.5$ being the Kolmogorov constant. The total energy density can be calculated by integrating over all wave numbers \cite{27B3}. \par

In this paper, we just focus our attention to the empirical decay law of the isotropic turbulence. Based on grid experiments in wind and water tunnels, the mean kinetic energy $\frac{1}{2}\overline {{\bf{u}}^2(t)}$ decays as $
 \overline {{\bf{u}}^2(t)}  \propto t^{-6/5} $ \cite{28B3,29B3}, as well as the theoretical treatment in \cite{27B3}.\par
Let us consider now the classical equation of motion for a viscous fluid  $ 
\partial_t (\rho_mu_i)+\partial_k\Pi_{ik}=0$
where $\rho_m$ is the mass density and $\Pi_{ik}$ the momentum
flux density tensor $
\Pi_{ik}=p\delta_{ik}+\rho_mu_iu_k-\mu
(\partial_ku_i+\partial_iu_k)$
where $\mu$ is the shear viscosity \cite{26B3}. Making use of the mean of this equation,
and notifying that $\bar{u}_i=0$ in homogeneous and isotropic
turbulence, we get
\begin{equation}
\bar{\Pi}_{xx}=\bar{\Pi}_{yy}=\bar{\Pi}_{zz} \equiv p_{\rm
eff}=p+\frac{2}{3}\rho_{\rm turb}, \quad p=\omega(t)\rho+\Lambda(t)\,\,,\label{ainamon8}
\end{equation}
where $\Lambda(t)$ is the variable cosmological constant, $\omega(t)$ a variable parameter of equation of state, and $p_{\rm eff}$ is the effective pressure which takes into account that the
thermodynamical pressure is augmented by a term
$(2/3)\rho_{\rm turb}$ associated with the turbulent energy
density $
\rho_{\rm turb}=\frac{1}{2}\rho_m {\overline{u_i u_i}} \equiv
\frac{1}{2}\rho_m \overline{{\bf{u}}^2}$, 
meaning that $\rho_{\rm turb}$ designates a mean quantity.
%%%%%%%%%%%%%%%%%%%%%%%%%%%%%%%%%%%%%%%%%%%%%%%%%%%%%%%%%%%%%%%%%%%%%%%%%%%%%%
%%%%%%%%%%%%%%%%%%%%%%%%%%%%%%%%%%%%%%%%%%%%%%%%%%%%%%%%%%%%%%%%%%%%%%%%%%%%%%
%%%%%%%%%%%%%%%%%%%%%%%%%%%%%%%%%%%%%%%%%%%%%%%%%%%%%%%%%%%%%%%%%%%%%%%%%%%%%%%%%
%%%%%%%%%%%%%%%%%%%%%%%%%%%%%%%%%%%%%%%%%%%%%%%%%%%%%%%%%%%%%%%%%%%%%%%%%%%%%%%%

\section{FRW equations using a fraction $f$ of the energy as turbulent energy}

Let us now consider this fluid in a cosmological setting. As
usual in turbulence theory we may start by decomposing the fluid
velocity $u_i$ into a mean component $U_i$ and a fluctuating
component $u_i'$, $u_i=U_i+u_i'$. On the other hand, in a comoving reference
frame $U_i=0$, so that we can simply replace $u_i'$ with $u_i$. Requiring only the turbulence  part of the cosmic fluid, far from the non-viscous (non-turbulence) part, one can from now adopt the notation for which the non-viscous terms are subscripted by zero. Thus Eq.~(\ref{ainamon8})
is rewritten as
\begin{equation}
p_{\rm eff}=p_0+\frac{2}{3}\rho_{\rm turb}. \label{E}
\end{equation}
The turbulence energy in the comoving frame can be view as a nonrelativistic quantity, and the total energy density can be written as
\begin{eqnarray}
\rho=\rho_0+\frac{2}{3}\rho_{turb}\;.
\end{eqnarray}
Let us look for the cosmological equations when a constant fraction $f$ of the energy exists in form of turbulence energy, and assume that the energy at initial time $t_{in}$ can be written as
\begin{eqnarray}
\rho(t_{in})=\rho_{0}\left(1+f\right)\;.\label{rhodet}
\end{eqnarray}
From this, we see that $f$ can be interpreted as the ratio between the turbulence energy and the total energy at $t_{in}$,
\begin{eqnarray}
f=\frac{\rho_{turb}}{\rho_0}|_{t=t_{in}}\;.
\end{eqnarray}
Since the rest mass is completely incorporated in the total energy, the condition $f<<1$ is required \cite{B3}. Therefore, for $t>t_{in}$ it is necessary $\rho_{turb}$ decaying with time as  $\rho_{turb}\propto t^{-6/5}$, according to the time evolution of the ordinary turbulence \cite{B3}. Thus, for $t\geq t_{in}$ the energy density can be written in the following form
\begin{eqnarray}
\rho(t)=\rho_0(t)\left[1+f\rho_1(t)\right]\;,\quad \rho_1(t)=\left(\frac{t_{in}}{t}\right)^{6/5}\;\;.
\end{eqnarray}
The choice of $\rho_1(t)$ seems quick suitable since for $t=t_{in}$, the relation (\ref{rhodet}) is recovered.\par
Let us now write the equation of state for the cosmic fluid as
\begin{eqnarray}
p_0(t)=\omega_0(t)\rho_0(t)+\Lambda_0(t)\,\,\,.\label{eosnv}
\end{eqnarray}
In the same way as (\ref{rhodet}), one can write the effective pressure as 
\begin{eqnarray}
p_{eff}(t)=\omega(t)\rho(t)+\Lambda(t), \;\;\omega(t)=\omega_0(t)\left[1+f\omega_1(t)\right],\;\;\;
\Lambda(t)=\Lambda_0(t)\left[1+f\Lambda_1(t)\right]\,\,.\label{eosseconddef}
\end{eqnarray}
Through the condition $f<<1$, one can rewrite the effective pressure as
\begin{eqnarray}
p_{eff}(t)=p_0(t)+\Big\{\omega_0(t)\rho_0(t)\left[\omega_1(t)+\rho_1(t)\right]+
\Lambda_0(t)\Lambda_1(t)\Big\}f\;.\label{pressureeff}
\end{eqnarray}
Analogously, the scale factor and the Hubble parameter can be expanded as 
\begin{eqnarray}
a(t)&=&a_0(t)\left[1+fa_1(t)\right]\,\,,\label{scale}\\
H(t)&=&H_0(t)\left[1+fH_1(t)\right]\,\,\,.\label{hubble}
\end{eqnarray}
The correction terms $\omega_1(t)$, $\Lambda_1(t)$, $a_1(t)$ and $H_1(t)$ are all for zero order in $f$. Using (\ref{hubble}) and deriving (\ref{scale}) one time with respect to $t$, one gets
\begin{eqnarray}
\dot{a}_1(t)=H_0(t)H_1(t)\;.
\end{eqnarray}
From the first equation of Friedmann, $3H^2=\rho$, where we set $\kappa=8\pi G=1$, $G$ being the Newtonian gravitational constant, one gets
\begin{eqnarray}
H_1(t)=\frac{1}{2}\rho_1(t)\,\,.
\end{eqnarray}
Hence, the Hubble parameter reads
\begin{eqnarray}
H(t)=H_0(t)\left[1+\frac{1}{2}f\left(\frac{t_{in}}{t}\right)^{6/5}\right]\,\,.\label{hubbleparameter}
\end{eqnarray}
Now, we also have to  determine $H_0(t)$, and in order to do this, we need to use the equation of continuity related to the non-viscous components as
\begin{eqnarray}
\dot{\rho}_0+3H_0\left(\rho_0+p_0\right)=0\,\,\,.\label{eqcontinuitynv}
\end{eqnarray}
Let us now consider for simplicity the case where both $\omega_0$ and $\Lambda_0(t)$ are  linear time dependent, i.e.,  $\omega_0(t)=\alpha t+\beta$ and $\Lambda_0(t)=\gamma t+\lambda$. Therefore, the equation (\ref{eqcontinuitynv}) becomes
\begin{eqnarray}
2\dot{H}_0+3H_0^2\left(1+\beta+\alpha t\right)+\gamma t+ \lambda=0\,\,,\label{eqdiff1}
\end{eqnarray} 
whose general solution reads
\begin{eqnarray}
H_0(t)=\frac{AH_0(t_{in})}{B\Big[e^{C(t+D)^2}\mbox{erf}(t)-e^{C(t_{in}+D)^2}\mbox{erf}(t_{in})\Big]+A}\,\,,
\end{eqnarray}
where the input constant are defined as
\begin{eqnarray}
C=\frac{\sqrt{3}}{4}\gamma,\quad D=\frac{\lambda}{\gamma},\quad A=\frac{C}{\alpha\sqrt{3}}\,,\quad B=\frac{1}{4}\sqrt{\frac{2\pi \sqrt{3}}{C}}\Bigg[\frac{\alpha C}{1+\beta}-\lambda\Bigg]\,.\nonumber
\end{eqnarray}
Therefore, the Hubble parameter (\ref{hubbleparameter}) becomes
\begin{eqnarray}
H(t)=\frac{AH_0(t_{in})}{B\Big[e^{C(t+D)^2}\mbox{erf}(t)-e^{C(t_{in}+D)^2}\mbox{erf}(t_{in})\Big]+A}\left[1+\frac{1}{2}f\left(\frac{t_{in}}{t}\right)^{6/5}\right]\,\,.
\label{Hubblegeneral}
\end{eqnarray}
An example illustrating the evolution of the Hubble parameter for some suitable values of the input parameters is presented in the Fig.$1$ at the upper-left side. On can see that the Hubble parameter is initially high and as the time evolves it goes to zero. This may correspond to an asymptomatically flat universe. The associated energy density reads
\begin{eqnarray}
\rho(t)=\frac{3A^2H_0^2(t_{in})}{\Big\{B\Big[e^{C(t+D)^2}\mbox{erf}(t)-e^{C(t_{in}+D)^2}\mbox{erf}(t_{in})\Big]+A\Big\}^2}\left[1+f\left(\frac{t_{in}}{t}\right)^{6/5}\right]\,\,.\label{densitygeneral}
\end{eqnarray}
Observe that in general, for small $t_{in}/t$ (or $t>>t_{in}$), the influence from the turbulence fades away for both the Hubble parameter and the energy density.\par
Let us now analyse some special cases where cosmological feature may be obtained.\par

\subsection{Analysing $\omega_0$ as linear function of time and $\Lambda_0$ a constant}
In this case, we have $\omega_0(t)=\alpha t+\beta$ and $\Lambda_0=\lambda$, meaning that $\gamma=0$. Then, by solving (\ref{eqdiff1}), the Hubble parameter and the energy density are found as
\begin{eqnarray}
H(t)&=&\frac{\sqrt{\lambda}H_0(t_{in})\bar{Z}(t)}{\sqrt{9\alpha H_0^2(t_{in})(t-t_{in})+\lambda\bar{Z}^2(t_{in})}}\left[1+\frac{1}{2}f\left(\frac{t_{in}}{t}\right)^{6/5}\right]\,\,,\\
\rho(t)&=&\frac{3\lambda H_0^2(t_{in})\bar{Z}^2(t)}{9\alpha H_0^2(t_{in})(t-t_{in})+\lambda\bar{Z}^2(t_{in})}\left[1+f\left(\frac{t_{in}}{t}\right)^{6/5}\right]\,\,,\\
\bar{Z}(t)&=&\frac{Z_{-\frac{1}{3}}\Big[\frac{1}{\alpha}\sqrt{\frac{\lambda}{3}}\left(\alpha t+\beta+1\right)^{\frac{3}{2}}\Big]}{Z_{\frac{2}{3}}\Big[\frac{1}{\alpha}\sqrt{\frac{\lambda}{3}}\left(\alpha t+\beta+1\right)^{\frac{3}{2}}\Big]}\,\,,\nonumber
\end{eqnarray}
where $Z_{\nu}=C_1J_{\nu}+C_2Y_{\nu}$ is the general solution of Bessel's differential equation, $C_1$ and $C_2$ being arbitrary constants. One can observe that for $t=-(\beta+1)/\alpha$ both the Hubble parameter and the energy density diverge simultaneously, corresponding to the singularity of type-I (the Big Rip).\par
%%%%%%%%%%%%%%%%%%%%%%%%%%%%%%%%%%%%%%%%%%%%%%%%%%%%%%%%%%%%%%%%%%%%%%%%
%%%%%%%%%%%%%%%%%%%%%%%%%%%%%%%%%%%%%%%%%%%%%%%%%%%%%%%%%%%%%%%%%%%%%%%%%%%%%
$\bullet$ {\bf Linear $\omega_0(t)$ and vanishing cosmological constant}\par
In this situation the cosmological constant is null, $\Lambda_0=0$, whilst $\omega_0(t)=\alpha t+\beta$. Thus, by solving (\ref{eqdiff1}), one gets  the  Hubble parameter and the energy density 
\begin{eqnarray}
H(t)=\frac{H_0(t_{in})}{H_0(t_{in})\left(t-t_{in}\right)\Big[\frac{3}{2}(1+\beta)+\frac{3\alpha}{4}\left(t+t_{in}\right)\Big]+1}\left[1+\frac{1}{2}f\left(\frac{t_{in}}{t}\right)^{6/5}\right]\,\,,\\
\rho(t)=3\frac{H_0^2(t_{in})}{\Big\{H_0(t_{in})\left(t-t_{in}\right)\Big[\frac{3}{2}(1+\beta)+\frac{3\alpha}{4}\left(t+t_{in}\right)\Big]+1\Big\}^2}\left[1+f\left(\frac{t_{in}}{t}\right)^{6/5}\right]\,\,.
\end{eqnarray}
These results generalize the ones found in \cite{B3}, which are recovered for $\alpha=0$.  The evolution the Hubble parameter is illustrated at the upper-right side of the Fig.$1$ for some suitable values of the input parameters. One can see that as the time evolves, the Hubble parameter decreases and goes toward zero, showing that the universe under consideration is asymptotically flat.\par
As in the vanishing turbulence case, finite time future singularity occurs at $t_s$ such that
\begin{eqnarray}
\alpha t_s^2+2\left(1+\beta\right)t_s-2\left(1+\beta\right)t_{in}-\alpha t^2_{in}+\frac{4}{3H_0(t_{in})}=0\,\,.\label{alvaro}
\end{eqnarray}
For $\alpha \neq 0$, one gets 
\begin{eqnarray}
t_s=-\frac{1+\beta}{\alpha}\pm\sqrt{\left(t_{in}+\frac{1+\beta}{\alpha}\right)^2-\frac{4}{3\alpha H_0(t_{in})}}\,\,. \label{eqdiffts}
\end{eqnarray}
Observe that, for $\alpha=0$, the singularity time reads\footnote{This value has to be directly used in equation (\ref{alvaro})}
\begin{eqnarray}
t_s=t_{in}+\frac{2}{3|1+\beta|H_0(t_{in})}\,\,.
\end{eqnarray}
This later is exactly the result found for the singularity time in \cite{B3}, where they used $\gamma$ at the place of $1+\beta$ and $\omega_0$(constant) at the place of $\beta$. Consequently all the results found in that paper should also be found here. Since the goal of this paper is to search for more general cases, let us reconsider the case where $\omega_0(t)$ is linear time dependent and the cosmological constant vanishes. One can now search for the correction term $a_1$ which is such that $\dot{a}_1=H_0\rho_1/2$. One gets
\begin{eqnarray}
a_1(t)=-\frac{1}{2\sqrt{B^2-4AC}}\left(\frac{2t_{in}}{t}\right)^{6/5}\Bigg\{f_1^{6/5}(t)\,_2F_1\left(\frac{6}{5},\frac{6}{5},\frac{11}{5},g_1(t)\right)-f_2^{6/5}(t)\,_2F_1\left(\frac{6}{5},\frac{6}{5},\frac{11}{5},g_2(t)\right)\Bigg\}\,\,,\label{expressiona1}
\end{eqnarray}
where the functions $f_i(t)$, and $g_i(t)$ are defined by
\begin{eqnarray}
f_1(t)=\frac{At}{B-\sqrt{B^2-4AC}+2At}\,\,,\quad f_2(t)=\frac{At}{B+\sqrt{B^2-4AC}+2At}\,\,,\nonumber\\
g_1(t)=\frac{B-\sqrt{B^2-4AC}}{B-\sqrt{B^2-4AC}+2At}\,\,,\quad g_2(t)=\frac{B+\sqrt{B^2-4AC}}{B+\sqrt{B^2-4AC}+2At}\,\,,\\
A=\frac{3\alpha}{4},\quad B=\frac{3}{2}\left(1+\beta\right),\quad C=\frac{1}{H_0(t_{in})}-\frac{3}{4}\alpha t^2_{in}-\frac{3}{2}\left(1+\beta\right)t_{in}\,\,,\nonumber
\end{eqnarray} 
whilst the scale factor $a_0(t)$ in the vanishing turbulence case reads
\begin{eqnarray}
a_0(t)=\frac{2}{\sqrt{4AC-B^2}}\arctan{\left(\frac{B+2At}{\sqrt{4AC-B^2}}\right)}\,\,.
\label{expressiona0}
\end{eqnarray}
Since we are leading with the case where the cosmological constant vanishes, the correction term $\Lambda_1$ is also null, ($\Lambda_1(t)=0$). We can now proceed to the calculation of the correction term $\omega_1(t)$. To this end, let us make use of the second generalized equation of  Friedmann 
\begin{eqnarray}
-3H^2(t)-2\dot{H}(t)=p_{eff}\,\,.\label{secondEqF}
\end{eqnarray}
The left hand side of (\ref{secondEqF}) can be calculated from (\ref{hubble}), and comparing the result with the right hand side of (\ref{pressureeff}), one gets
\begin{eqnarray}
\omega_1(t)&=&-\rho_1(t)-\frac{1}{\omega_0(t)\rho_0(t)}\Bigg\{\left[\dot{H}_0(t)+3H_0^2(t)\right]\rho_1(t)+H_0(t)
\dot{\rho}_1(t)\Bigg\}\,\,,\nonumber\\
&=&\frac{(16A-15\alpha)t^2+(B-15-15\beta)t+6C}{15t(\alpha t+\beta)}\left(\frac{t_{in}}{t}\right)^{6/5}\,\,\,.\label{correctionomega}
\end{eqnarray}
Therefore, with an inhomogeneous equation of state of the form (\ref{eosseconddef}), with vanishing cosmological constant, the correction term $\omega_1(t)$ associated to $\omega_0(t)$ is given by (\ref{correctionomega}). Observe that at singularity time, $t_s>t_{in}$, this correction term becomes inefficient such that the turbulence effect in $\omega(t)$ falls down and this latter reduces to $\omega_0(t)$.

\subsection{Analysing $\Lambda_0$ as linear function of time and $\omega_0$ a constant}
In this case we consider $\alpha=0$ such that $\omega_0=\beta$ and $\Lambda_0(t)=\gamma t+\lambda$. Thereby, the general solution of (\ref{eqdiff1}) becomes
\begin{eqnarray}
H(\xi)&=&H_0(\xi_{in})\frac{Y(\xi)}{Y(\xi_{in})}\Bigg[1+\frac{1}{2}f\left(\frac{4\xi_{in}-3\lambda}{4\xi-3\lambda}\right)^{6/5}\Bigg]\,\,,\label{hubblexi}\\
Y(\xi)&=&\frac{2\left(3\xi^3-2\sqrt{\sigma\xi}\right)}{3\xi^3-2\xi\sqrt{\sigma\xi}+\sigma}\,,\quad \xi=\frac{3}{4}\left(\gamma t+\lambda\right)\,\quad \sigma= \frac{4}{3\gamma}\,\,.\nonumber
\end{eqnarray}
Here, it is important to note that this solution makes sense only if $\beta \neq -1$, otherwise, we fall in the case where the Hubble parameter is a polynomial function of the cosmic time. But we will see this later. An numerical example illustrative of the evolution of the Hubble parameter is presented at the lower-left side in the Fig.$1$. We see from this that the Hubble parameter initially increases and  decreases at the late times. This situation may be view as an initially accelerated universe, which later enters in a decelerated phase.\par
Further more we note that in this case of constant $\omega_0$ and linear $\Lambda_0$, the energy density reads
\begin{eqnarray}
\rho(\xi)=3H^2_0(\xi_{in})\frac{Y^2(\xi)}{Y^2(\xi_{in})}\Bigg[1+f\left(\frac{4\xi_{in}-3\lambda}{4\xi-3\lambda}\right)^{6/5}\Bigg].\label{edensityxi}
\end{eqnarray}
Since the parameter $\beta$ is non-null, one can easily observe that both energy density and Hubble parameter diverge only for complex value of $\xi$, consequently, for complex value of the cosmic time. Therefore, one can conclude that this case is interesting since the big rip cannot appear for finite real time. Note also that since the $\omega_0$ is constant, its correction term $\omega_1$ vanishes. Thus, one has now to search for the correction term $\Lambda_1(t)$ corresponding to $\Lambda_0(t)$. As we have performed in the previous case, now looking for $\Lambda_1(t)$, one gets
\begin{eqnarray}
\Lambda_1(t)=-\frac{1}{\Lambda_0(t)}\left\{\omega_0(t)\rho_0(t)\rho_1(t)
+6H_0^2(t)H_1(t)+\dot{H}_0(t)H_1(t)+H_0(t)
\dot{H}_1(t)\right\}\,\,,
\end{eqnarray}
which, in terms of $\xi$, yields
\begin{eqnarray}
\Lambda_1=-\frac{6H_0(\xi_{in})}{4\xi Y(\xi_{in})}\Bigg\{3\frac{H_0(\xi_{in})}{Y(\xi_{in})}Y^2(\xi)+\frac{\gamma}{8}Y'(\xi)-\frac{3\gamma}{5(4\xi-3\lambda)}Y(\xi)\Bigg\}\left(\frac{4\xi_{in}-3\lambda}{4\xi-3\lambda}\right)^{6/5}\,\,,\\
Y'(\xi)=\frac{4\sqrt{\sigma\xi}\left(-9\xi^3+3\xi^2+9\xi+\sigma+\sqrt{\sigma}\right)
-6\sigma\xi\left(\xi^2-3\xi+2\right)-\sigma\sqrt{\sigma/\xi}}{\left(3\xi^3-2\xi\sqrt{\sigma\xi}+\sigma\right)^2}\,\,.\nonumber
\end{eqnarray}
Note that the sub-case where the parameter $\omega_0$ vanishes whilst $\Lambda_0(t)$ is still linear function of time can be easily found from this later by setting $\beta=0$. Is important to recall that the results are physically the same, because this is just a consequence of rescaling the parameter of Hubble by a multiplicative constant (leaving $\beta+1$ to $1$) in the equation (\ref{eqdiff1}). Other interesting case to be mentioned is that in which both parameters $\omega_0$ and $\Lambda$ are constants. In this case, one has $\alpha=\gamma=0$, $\omega_0=\beta$ and $\Lambda_0=\lambda$. Expressions of the Hubble parameter and the energy density may be obtained by making use of the above assumptions in (\ref{hubblexi})-(\ref{edensityxi}). In this subsection, one can conclude that for a constant $\omega_0$ and linear time dependent cosmological constant, future singularity can never occur, at least for real cosmic time.

\subsection{Treating $\omega_0$ as an oscillating function, with a vanishing cosmological constant}
Here, instead of using $\omega_0(t)$ as linear function of time, we assume that it is an oscillating function of time. Let us consider $\omega_0(t)=-1+l_0\cos(b\,t)$. By making use of this later and vanishing the cosmological constant, one gets from the equation (\ref{eqdiff1})
\begin{eqnarray}
H_0(t)=\frac{2b}{3\left(l_1+l_0\sin(b\,t)\right)}\,\,,\label{1hubbleosci}
\end{eqnarray}
and the Hubble parameter within turbulence reads
\begin{eqnarray}
H(t)=\frac{bH_0(t_{in})}{b+3l_0H_0(t_{in})\cos\left[b(t+t_{in})/2\right]
\sin\left[b(t-t_{in})/2\right]}\left[1+\frac{1}{2}f\left(\frac{t_{in}}{t}\right)^{6/5}\right]\,\,,
\end{eqnarray}
where $l_1$ is an integration constant. We present an numerical illustrative evolution of the Hubble parameter at the lower-right side of Fig.$1$. As expected, this figure shows and oscillating evolution of the Hubble parameter, corresponding to an alternative universe, i.e., an universe always transiting from a decelerated phase to an accelerated phase and vice-versa. The associated energy density reads
\begin{eqnarray}
\rho(t)=\frac{3b^2H_0^2(t_{in})}{\Big\{b+3l_0H_0(t_{in})\cos\left[b(t+t_{in})/2\right]
\sin\left[b(t-t_{in})/2\right]\Big\}^2}\left[1+f\left(\frac{t_{in}}{t}\right)^{6/5}\right]\,\,.
\end{eqnarray}
It is also obvious here that for $t>>t_{in}$, the turbulence effect fades away both for the Hubble parameter and the energy density. In order to analyse very clearly the occurrence of finite time singularity, let us reconsider the expression (\ref{1hubbleosci}). Note that if $|l_1|<l_0$, the denominator can be zero, which implies a cosmological singularity. However, if $|l_1|>l_0$, singularity cannot occur.
It appears that both BiG Rip and Little Rip may appear in dark energy within turbulence consideration. This kind of results has been found recently by Brevik et al \cite{B4} but not within oscillating inhomogeneous EoS.
%This is an important result, since, besides of leading with a phantom region of the universe, it does not end with a finite time singularity (Big Rip for instance). This is a similar behaviour observed in little rip cosmology. Recently, Brevik et al shown that little behaviour for dark energy can be found \cite{B4}.

\par
The correction term $\omega_1(t)$ corresponding to $\omega_0(t)$, can be found in the same way as the previous cases by
\begin{eqnarray}
\omega_1(t)&=&-\frac{2}{\omega_0(t)\rho_0(t)}\left[3H^2_0(t)H_1(t)+
\dot{H}_0(t)H_1(t)+H_0(t)\dot{H}_1(t)\right]
-\rho_1(t)\nonumber\\
&=&\left\{\frac{1}{1-l_0\cos(bt)}\left[1-\frac{2}{5t}-\frac{l_0}{2}\cos(bt_{in})\right]-1\right\}\left(\frac{t_{in}}{t}\right)^{6/5}\,\,,
\end{eqnarray}
which fades away for $t>>t_{in}$.

%%%%%%%%%%%%%%%%%%%%%%%%%%%%%%%%%%%%%%%%%%%%%%%%%%%%%%%%%%%%%%%%%%%%%%%%%%%%%%
%%%%%%%%%%%%%%%%%%%%%%%%%%%%%%%%%%%%%%%%%%%%%%%%%%%%%%%%%%%%%%%%%%%%%%%%%%%%%%%
%%%%%%%%%%%%%%%%%%%%%%%%%%%%%%%%%%%%%%%%%%%%%%%%%%%%%%%%%%%%%%%%%%%%%%%%%%%%%%%
\section{Turbulence as additive dark energy and stability analysis}
In this section, instead of using the perturbative approach, we assume the total dark energy density $\rho_{td}$ as the sum of the usual dark energy $\rho_{d}$ and the turbulence energy $\rho_{turb}$
\begin{eqnarray}
\rho_{td}=\rho_d+\rho_{turb}\,\,,
\end{eqnarray}
and the first equation of Friedmann is written as 
\begin{eqnarray}
3H^2=\rho_{d}+\rho_{turb}+\rho_{rad}\,\,,
\end{eqnarray}
where $\rho_{rad}$ characterises the energy density of  the radiation. Treating the turbulence, the dark energy and the radiation as three component fluid of the universe, the corresponding continuity equations read
\begin{eqnarray}
\dot{\rho}_d+3H\left(\rho_d+p_d\right)&=&Q_1\,\,\,,\\
\dot{\rho}_{rad}+3H\left(\rho_{rad}+p_{rad}\right)&=&Q_2\,\,\,,\\
\dot{\rho}_{turb}+3H\left(\rho_{turb}+p_{turb}\right)&=&Q_3\,\,\,,
\end{eqnarray}
such that  $Q_1+Q_2+Q_3=0$. We also define dimensionless density parameters as
\begin{eqnarray}
x\equiv \frac{\rho_d}{3H^2}\,\,,\quad y\equiv \frac{\rho_{rad}}{3H^2}\,\,,\quad z\equiv \frac{\rho_{turb}}{3H^2}\,\,\,\,.
\end{eqnarray}
The continuity equations in dimensionless variables reduce to
\begin{eqnarray}
\left\{\begin{array}{lll}
\frac{dx}{dN}&=&3x\left[x\left(1+\omega_d\right)+y\left(1+\omega_{rad}\right)
+z\left(1+\omega_{turb}\right)\right]-3x\left(1+\omega_d\right)+\frac{Q_1}{3H^3}\,\,,\\
\frac{dy}{dN}&=&3y\left[x\left(1+\omega_d\right)+y\left(1+\omega_{rad}\right)
+z\left(1+\omega_{turb}\right)\right]-3y\left(1+\omega_{rad}\right)+\frac{Q_2}{3H^3}\,\,,\label{firstsystem}\\
\frac{dz}{dN}&=&3z\left[x\left(1+\omega_d\right)+y\left(1+\omega_{rad}\right)
+z\left(1+\omega_{turb}\right)\right]-3z\left(1+\omega_{turb}\right)+\frac{Q_3}{3H^3}\,\,,\end{array}\right.
\end{eqnarray}
where $N\equiv \ln{a}$ is the so-called e-folding parameter. In general the interaction terms $Q_{i}$, $i=1,2,3$, are functions of the Hubble parameter energy density $\rho_j$. One obtains the critical points by equating the equations in (\ref{firstsystem}) to zero. Next, we will perturb the equations up to first order around the critical points and analyse their stability. Only the stable critical point will be taken into account, i.e, the critical points for which all the eigenvalues of Jacobian matrix are negative. These points are interpreted as attractor solutions of the dynamical system.\par
In this paper we will be restricted to pure physical situation. Because of the turbulence, the kinetic energy changes into heat and the heat then becomes radiation. Therefore, the interaction essential occurs between the turbulence and the radiation. Consequently, we may set the interaction term $Q_1$ to zero. As well known, we can make use of the value of EoS of radiation  ($\omega_{rad}=1/3$). Since the total content of the universe has to conserve, one gets $Q_2=-Q_3=Q$. In such a situation, since dark energy is assumed to not interact with the radiation, or the turbulence, one can  consider its related parameter of equation of motion $\omega_d$ as constant and more precisely about $-1$. Therefore, the dynamics of the system essentially depends on the turbulence and the radiation. Hence, the system (\ref{firstsystem}) reduces to
\begin{eqnarray}\label{secondsystem}
\left\{\begin{array}{ll}
\frac{dy}{dN}=-4y+\frac{Q}{3H^3}\,\,,\\
\frac{dz}{dN}=-3z\left(1+\omega_{turb}\right)-\frac{Q}{3H^3}\,\,\,
\end{array}\right.
\end{eqnarray}
We assume the interaction $Q$ between the turbulence and the radiation as a quantity proportional to  $\rho_{rad}\rho_{turb}/H$, i.e,  $Q=q\rho_{rad}\rho_{turb}/3H$, where $q$ is a real parameter. Thus, the system (\ref{secondsystem}) becomes
\begin{eqnarray}
\left\{\begin{array}{ll}
\frac{dy}{dN}=-4y+qyz\,\,,\\
\frac{dz}{dN}=-3z\left(1+\omega_{turb}\right)-qyz\,\,.
\end{array}\right.
\end{eqnarray}
As inhomogeneous equation of state, we assume $\omega_{turb}=k_1H^{-2}\rho_{turb}/3+k_2$, i.e, $\omega_{turb}=k_1z+k_2$, where $k_1$ and $k_2$ are two negative real constants. This choice of the sign of $k_1$ and $k_2$ is purely cosmological since as the turbulence contributes as a type of dark fluid, its pressure has to be negative  in order to insure the acceleration of the universe and a necessary condition for this is assuming $k_1$ and $k_2$ as negative constants. The system presents two realistic critical points: a)$ (y_c,z_c)=\left[0,-(1+k_2)/k_1\right]$ and b) $\left[-3(q+qk_2+4k_1)/q^2,4/q\right]$.  The first critical point correspond to the situation where the kinetic energy of the turbulence is still conserved, i.e, there is any change of the kinetic energy into heat. However, the second critical point is related to the situation where there is a real dynamics of the system. In this case,  part of kinetic energy is lost, yielding radiation. We will focus our attention to this second critical point and analyse the stability of the system about it. To this end, we will consider small perturbations around this critical point and write any point $(y,z)$ of the system as $\left(y+\delta y, z+\delta z\right)$. Thereby, the dynamical system becomes
\begin{eqnarray}\label{lastsystem}\left\{\begin{array}{ll}
\frac{d\delta y}{dN}=-3\left(1+k_2+4\frac{k_1}{q}\right)\delta z\,\,,\\
\frac{d\delta z}{dN}=-4\delta y-12\frac{k_1}{q}\delta z
\end{array}\right.
\end{eqnarray}
By denoting $\mathcal{M}$ the matrix associated to the above system, one can find the eigenvalues  by solving the equation $det\left(\mathcal{M}-\tilde{\lambda}I\right)=0$. The solutions of this equations read
\begin{eqnarray}
\tilde{\lambda}_{1,2}=-\frac{6k_1}{q}\pm\sqrt{\frac{36k_1^2}{q^2}+12\left(1+k_2+\frac{4k_1}{q}\right)}\,\,.
\end{eqnarray}
It can easily been seen that the sign of the eigenvalues strongly depend on that of the interaction parameter $q$. One can observe two general cases: $q<0$ and $q>0$. For $q<0$ and $k_2<-1-4k_1/q$, one gets $-6k_1/q<0$ and both $\tilde{\lambda}_{1}$ and $\tilde{\lambda}_2$ are negative. In this case, the critical point is an attractor and the system may be stable. Still for $q<0$, if $-1-4k_1/q<k_2<0$, $\tilde{\lambda}_1>0$ and $\tilde{\lambda}_2<0$ and the critical point is called saddle point and the system cannot be stable. For $q>0$, $-6k_1/q>0$, always one gets $\tilde{\lambda}_1\tilde{\lambda}_2<0$, or $\tilde{\lambda}_1>0$ and $\tilde{\lambda}_2>0$. In such situations, the system cannot be stable.  \par
We clearly see that the stability of the system is realized only when $k_1/q>0$ and $k_2<-1-4k_1/q$. By the use of these conditions, we from (\ref{lastsystem}) that as the universe expands, i.e, for increasing $N$ (increasing scale factor), the perturbation of the turbulence and radiation energy densities decreases.

\newpage

\begin{center}
\begin{figure}[t]
\begin{minipage}[t]{0.45\linewidth}
\includegraphics[width=\linewidth]{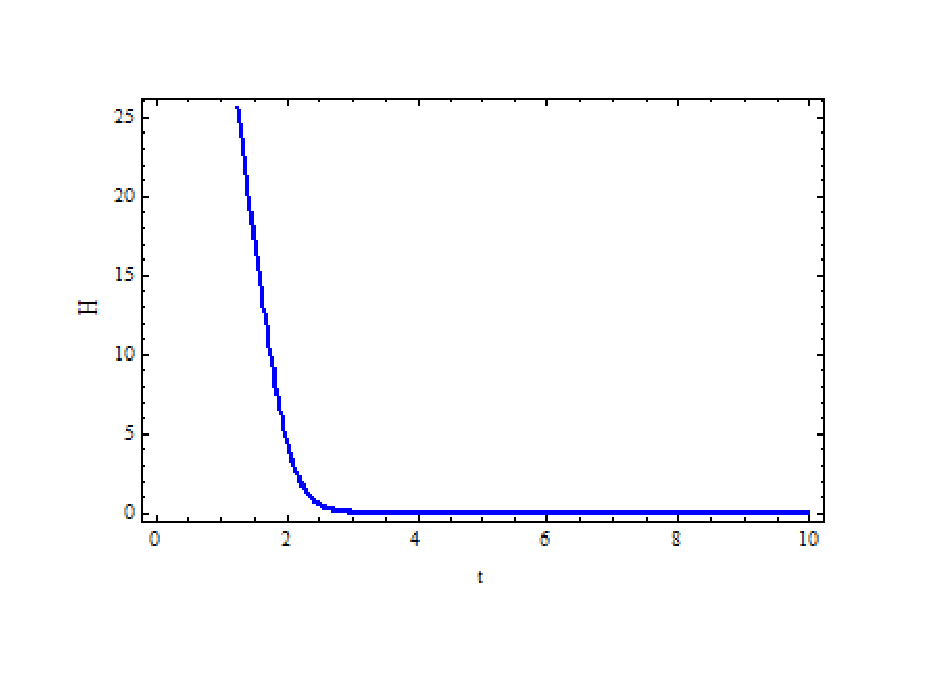}
\end{minipage} \hfill
\begin{minipage}[t]{0.45\linewidth}
\includegraphics[width=\linewidth]{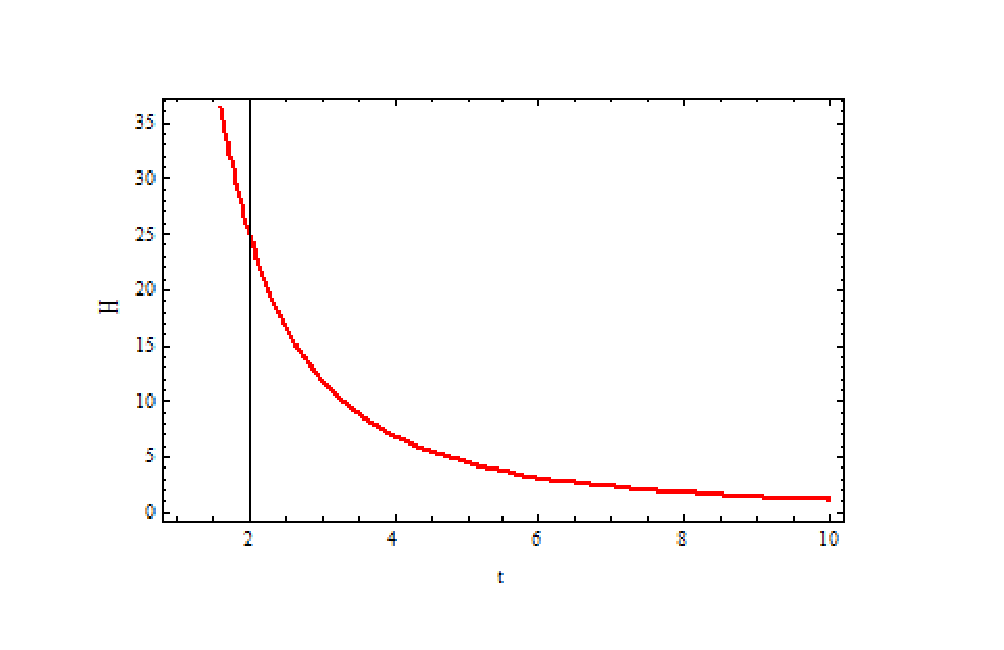}
\end{minipage} \hfill
\begin{minipage}[t]{0.45\linewidth}
\includegraphics[width=\linewidth]{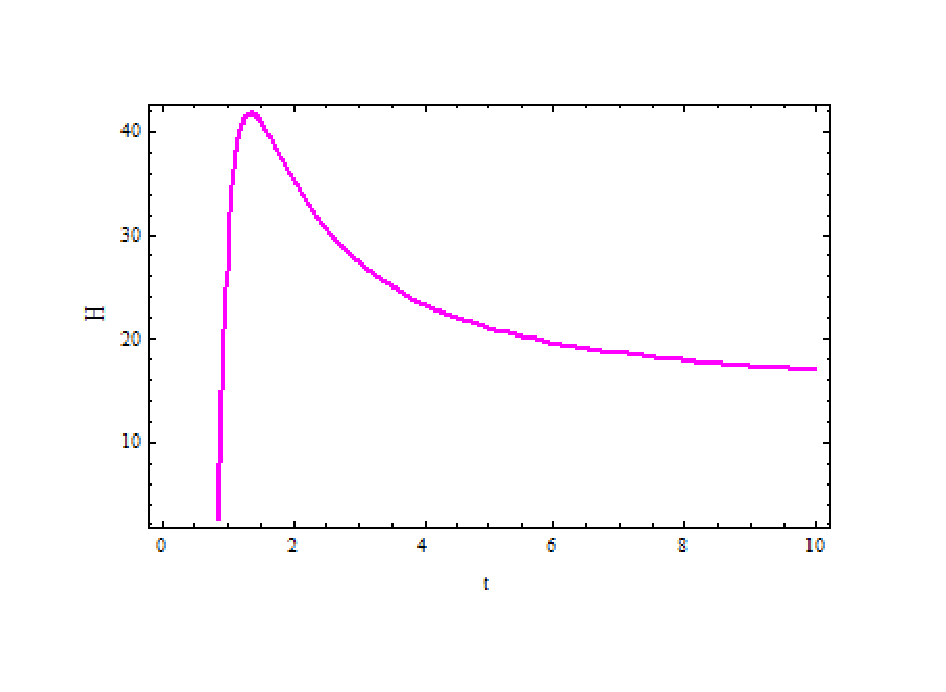}
\end{minipage} \hfill
\begin{minipage}[t]{0.45\linewidth}
\includegraphics[width=\linewidth]{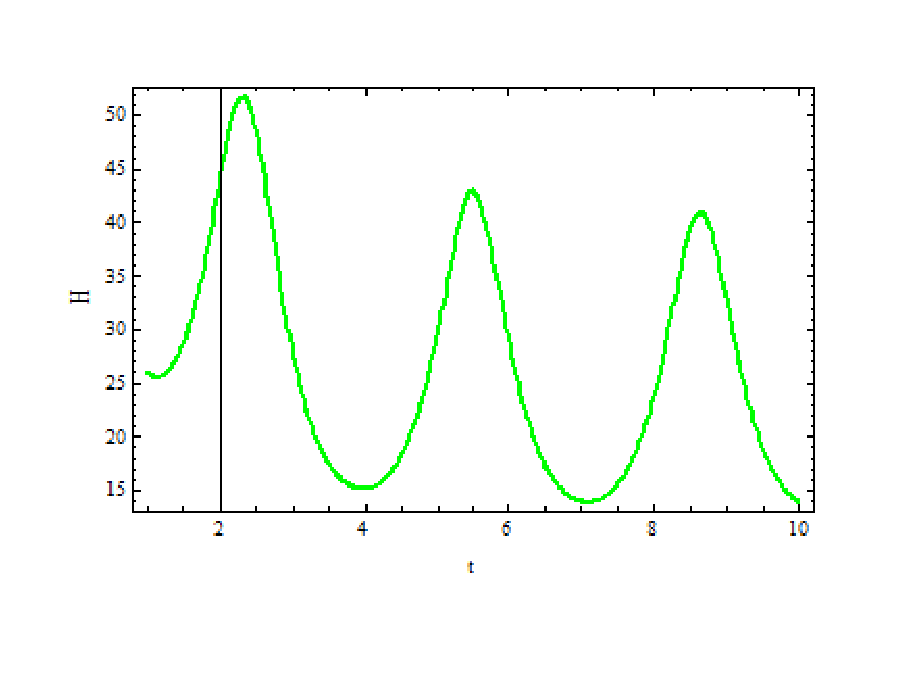}
\end{minipage} \hfill
\caption{{\protect\footnotesize These figures show the evolution of the Hubble parameter for various assumptions assumed in the manuscript and also for some appropriate values of the input parameters. The blue (upper left) corresponds to the case where both the parameter of EoS and the  cosmological constant are time linear dependent (Eq.~(18)). The red (upper right) corresponds to the case of time linear dependent EoS and vanishing cosmological constant (Eq.~(22)).  The magenta (lower left) characterizes the evolution of the Hubble parameter for 
linear cosmological constant and constant EoS while the green (lower right) one corresponds to the case oscillating case governed by the parameter of EoS and a vanishing cosmological constant.}}
\label{}
\end{figure}

\end{center}

%Since the current universe is expanding, the Hubble parameter is always positive and also the energy densities are positive. Therefore, from the relation $Q=q\rho_{rad}\rho_{turb}/(3H^3)$, it appear that $Q$ is also positive. By looking for the system (\ref{secondsystem}) wee see that with respect to the interaction term, as the universe expands the turbulence energy density decreases. This corresponds to a loss in kinetic energy, which is  converting into heat, yielding radiation. At the same, one observes that the energy density of radiation creases as the universe expands. On the other hand, it can be observe for $q<0$ ($Q<0$), the system is unstable and this corresponds to the situation on which the energy density of the radiation decreases while that of the turbulence is growing. Therefore, we conclude that the stability of the system is realized only when the energy density of the turbulence decreases, converting in heat characterises by the radiation.

\section{Conclusion}
In this paper, we investigate the effects of turbulence in dark fluid universe at late accelerated expanding universe. We focussed our attention to the model of linear inhomogeneous EoS, with linear and oscillating dependence on time. We included turbulence effect  by the  addition of a constant fraction to the laminar ordinary energy density. We then performed a perturbative analyse  checking the possible appearance of future singularity. The obtained results generalize that of Brevik et al \cite{B3}. Moreover, for each considered  expression for the parameters $\omega(t)$ and $\Lambda(t)$, we determined their respective correction terms.\par
In the second step, instead of considering the perturbative approach, we considered the dark energy contribution as a sum of a laminar part of dark energy and the turbulence part. Then, we study the stability of the system, considering that the dynamics of the system essentially depends on the turbulence and radiation energy densities. The results shows that the stability of the system is obtained when the turbulence energy density decreases as the universe expands, while radiation energy density of radiation grows. Therefore, when turbulence contribution is taking into account in the total dark fluid, according to the statement that kinetic energy changes into heat, the stability is always realized. Hence, precisely in phantom region, due to this probable stability, one may conclude that turbulence may avoid future finite time singularities, and this when all the input parameters $q$, $k_1$ and $k_2$ are negative and  satisfy the relations $k_1/q>0$ and  $k_2<-1-4k_1/q$.

\vspace{0.5cm}
{\bf Acknowledgement:} M. J. S. Houndjo thanks  ENS-Natitingou for partial financial support. M. E. Rodrigues thanks UFES for the hospitality during the elaboration of this work and also thanks CNPq and UFPA for financial support.


\begin{thebibliography}{90}

\bibitem{1B1} E. Copeland, M. Sami, S. Tsujikawa, Int. J. Mod. Phys. D
{\bf 15}, 1753 (2006).

\bibitem{1B12} T. Padmanabhan, Phys. Rep. {\bf 380}, 235 (2003).

\bibitem{1B13} L. Perivolaropoulos, astro-ph/0601014.

\bibitem{4B1} I.Y. Arefeva, I. Volovich, Theor.Math.Phys {\bf 155}, 503-511 (2008), hep-th/0612098.

\bibitem{1B3} R. R. Caldwell, M. Kamionkowski and N. N. Weinberg, Phys. Rev. Lett. {\bf 91}, 071301 (2003) [arXiv:astro-ph/0302506].

\bibitem{1B31} B. McInnes, JHEP 0208 (2002) 029 [arXiv:hep-th/0112066].

\bibitem{1B32}
S. Nojiri and S. D. Odintsov, Phys. Lett. B 562, 147 (2003) [arXiv:hep-th/0303117].

\bibitem{1B33}
E. Elizalde, S. Nojiri and S. D. Odintsov, Phys. Rev. D 70, 043539 (2004)[arXiv:hep-th/0405034].

\bibitem{1B34}
V. Faraoni, Int. J. Mod. Phys. D 11, 471 (2002) [arXiv:astro-ph/0110067].

\bibitem{1B35}
P. F. Gonzalez-Diaz, Phys. Lett. B 586, 1 (2004) [arXiv:astro-ph/0312579].

\bibitem{1B36}
C. Csaki, N. Kaloper and J. Terning, Annals Phys. 317, 410 (2005) [arXiv:astro-ph/0409596].

\bibitem{1B37}
 P. X. Wu and H. W. Yu,
Nucl. Phys. B 727, 355 (2005) [arXiv:astro-ph/0407424].

\bibitem{1B38}
 S. Nesseris and L. Perivolaropoulos, Phys. Rev. D {\bf 70}, 123529
(2004) [arXiv:astro-ph/0410309].

\bibitem{1B39}
 M. Sami and A. Toporensky, Mod. Phys. Lett. A {\bf 19}, 1509 (2004) [arXiv:gr-qc/0312009].

\bibitem{1B40} 
H. Stefancic, Phys. Lett. B {\bf 586}, 5 (2004) [arXiv:astro-ph/0310904].

\bibitem{1B41}
L. P. Chimento and R. Lazkoz, Mod. Phys. Lett. A {\bf 19}, 2479 (2004) [arXiv:gr-qc/0405020].

\bibitem{1B42}
E. Babichev, V. Dokuchaev and Yu. Eroshenko, Class. Quant. Grav. {\bf 22}, 143 (2005) [arXiv:astro-ph/0407190].

\bibitem{1B43}
X. F. Zhang, H. Li, Y. S. Piao and X. M. Zhang, Mod. Phys. Lett. A {\bf 21}, 231 (2006) [arXiv:astro-ph/0501652].

\bibitem{1B44}
E. Elizalde, S. Nojiri, S. D. Odintsov and P. Wang, Phys. Rev. D {\bf 71}, 103504 (2005) [arXiv:hep-th/0502082].

\bibitem{1B45}
M. P. Dabrowski and T. Stachowiak, Annals Phys. {\bf 321}, 771 (2006) [arXiv:hep-th/0411199].

\bibitem{1B46}
I. Y. Aref\'eva, A. S. Koshelev and S. Y. Vernov, Phys. Rev. D {\bf 72}, 064017 (2005) [arXiv:astro-ph/0507067].

\bibitem{1B47}
E. M. Barbaoza and N. A. Lemos, Gen. Rel. Grav. {\bf 38}, 1609 (2006) [arXiv:gr-qc/0606084].

\bibitem{19B3} S. Nojiri, S. D. Odintsov and S. Tsujikawa, Phys. Rev. D {\bf 71}, 063004 (2005) [arXiv:hep-th/0501025].

\bibitem{B3} I. Brevik, O. Gorbunova, S. Nojiri and S. D. Odintsov, Eur. Phys. J. C. {\bf 71}, 1629 (2011); arXiv:1011.6255v2 [hep-th].

\bibitem{5B2} E. Copeland, M. Sami, S. Tsujikawa, Int. J. Mod. Phys. D
{\bf 15}, 1753 (2006) [hep-th/0603057].

\bibitem{6B2}  T. Padmanabhan, Phys. Rep. {\bf 380}, 235 (2003).

\bibitem{7B2} V. Cardone, C. Tortora, A. Troisi, S. Capozziello, Phys.
Rev. D {\bf 73}, 043 508 (2006).

\bibitem{B221}
S. Nojiri, S.D. Odintsov, Phys. Rev. D {\bf 70}, 103 522 (2004)
[hep-th/0408170].

\bibitem{B1} I. Brevik, E. Elizalde, O. Gorbunova and A. V. Timoshkin, Euro. Phys. J. C. {\bf 52}, 223-228 (2007).


\bibitem{B2} I. Brevik, O. Gorbunova and A. V. Timoshkin, Euro. Phys. J. C. {\bf 51}, 179-183 (2007).


\bibitem{steph} S. J. M. Houndjo, Euro. Phys. Lett. {\bf 94}, 49001 (2011)  arXiv:1103.3006 [astro-ph.CO].


\bibitem{B222}  H. Stefancic, Phys. Rev. D {\bf 71}, 084 024 (2005).

\bibitem{B223} S. Capozziello, V. Cardone, E. Elizalde, S. Nojiri,
S.D. Odintsov, Phys. Rev. D {\bf 73}, 043 512 (2006) [astro-ph/
0508350].

\bibitem{B224} I. Brevik, O. Gorbunova, Y.A. Shaido, Int. J. Mod. Phys.
D {\bf 14}, 1899 (2005).


\bibitem{B225} J. Ren, X. Meng, Phys. Lett. B {\bf 633}, 1 (2006) [astro-ph/
0511163].

\bibitem{B226} I. Brevik, Int. J. Mod. Phys. D 15, 767 (2006) [gr-qc/
0601100].

\bibitem{14B2} S. Nojiri, S. D. Odintsov, hep-th/0702031.


\bibitem{25B3} S. Panchev, Random Functions and Turbulence (Pergamon Press, Oxford, 1971).

\bibitem{27B3} I. Brevik, Z. Angew. Math. Mech. {\bf 72}, 145 (1992).

\bibitem{26B3} L. D. Landau and E. M. Lifshitz, Fluid Mechanics, 2nd ed. (Pergamon Press, Oxford, 1987).

\bibitem{28B3}  K. R. Sreenivasan, S. Tavoularis and R. Henry, J. Fluid Mech. {\bf 100}, 597 (1980).

\bibitem{29B3} G. Rosen, J. Fluid Mech. {\bf 180}, 87 (1987).

\bibitem{B4} I. Brevik, R. Myrzakulov, S. Nojiri and S. D. Odintsov, Physical Review D {\bf 86}, 063007 (2012).

















 
\end{thebibliography}
\end{document}